# A Neural Ordinary Differential Equation Model for Visualizing Deep Neural Network Behaviors in Multi-Parametric MRI based Glioma Segmentation

Zhenyu Yang[1], Zongsheng Hu[2], Hangjie Ji[3], Kyle Lafata[1,4,5], Eugene Vaios[1], Scott Floyd[1], Fang-Fang Yin[1,2], Chunhao Wang[1*]

[1]Deparment of Radiation Oncology, Duke University, Durham, NC 27710
[2]Medical Physics Graduate Program, Duke Kunshan University, Kunshan, Jiangsu, China 215316
[3]Department of Mathematics, North Carolina State University, Raleigh, NC 27695
[4]Department of Radiology, Duke University, Durham, NC, 27710
[5]Department of Electrical and Computer Engineering, Duke University, Durham, NC, 27710

**Short Running Title: Neural ODE for glioma segmentation**

*Corresponding authors:
Chunhao Wang, Ph.D.
Box 3295, Department of Radiation Oncology
Duke University Medical Center
Durham, NC, 27710, United States
E-mail: chunhao.wang@duke.edu

Funding Statement: This work is partially supported by NIH CA014236

Conflict of Interest: None associated with this work

Abstract

*Purpose:*

To develop a neural ordinary differential equation (ODE) model for visualizing deep neural network behavior during multi-parametric MRI (mp-MRI) based glioma segmentation as a method to enhance deep learning explainability.

*Methods:*

By hypothesizing that deep feature extraction can be modeled as a spatiotemporally continuous process, we implemented a novel deep learning model, neural ODE, in which deep feature extraction was governed by an ODE parameterized by a neural network. The dynamics of 1) MR images after interactions with the deep neural network and 2) segmentation formation can thus be visualized after solving the ODE. An accumulative contribution curve (ACC) was designed to quantitatively evaluate each MR image's utilization by the deep neural network towards the final segmentation results.

The proposed neural ODE model was demonstrated using 369 glioma patients with a 4-modality mp-MRI protocol: T1, contrast-enhanced T1 (T1-Ce), T2, and FLAIR. Three neural ODE models were trained to segment enhancing tumor (ET), tumor core (TC), and whole tumor (WT), respectively. The key MR modalities with significant utilization by deep neural networks were identified based on ACC analysis. Segmentation results by deep neural networks using only the key MR modalities were compared to the ones using all 4 MR modalities in terms of Dice coefficient, accuracy, sensitivity, and specificity.

*Results:*

All neural ODE models successfully illustrated image dynamics as expected. ACC analysis identified T1-Ce as the only key modality in ET and TC segmentations, while both FLAIR and T2 were key modalities in WT segmentation. Compared to the U-Net results using all 4 MR modalities, Dice coefficient of ET (0.784→0.775), TC (0.760→0.758), and WT (0.841→0.837) using the key modalities only had minimal differences without significance. Accuracy, sensitivity, and specificity results demonstrated the same patterns.

*Conclusion:*

The neural ODE model offers a new tool for optimizing the deep learning model inputs with enhanced explainability. The presented methodology can be generalized to other medical image-related deep learning applications.

**Keywords:** Neural ODE, deep learning, visualization, explainability, glioma segmentation

## 1. Introduction

Gliomas, including low-grade glioblastoma (LGG) and high-grade glioblastoma (HGG), are the most severe brain tumor types with extremely high mortality and morbidity [1-5]. Radiological analysis, particularly magnetic resonance imaging (MRI), has become standard-of-care in glioma management [2, 4, 6, 7]. It is important to acquire accurate delineation of glioma boundaries with potential sub-region segmentations in personalized glioma treatment regimens for improved survival time [8, 9]. Multi-parametric MRI protocols that involve multiple anatomical and functional sequences can provide comprehensive and possible complementary information of spatial disease distribution [7, 10-13]; therefore, multi-parametric MRI has been extensively studied for glioma segmentation.

Currently, glioma segmentation is mainly done through manual delineations by experienced neuroradiologists [8, 14]. Such manual segmentation could be very time-consuming, and results could be subject to large intra- and inter-rater variability [15, 16]. Many efforts have thus been made for semi- or fully-automatic glioma segmentation for both efficiency and potential accuracy improvements [17]. Earlier efforts include $1^{st}$-order intensity thresholding [18] and region-based approach [19] for regional segmentation, but the reported performance was limited [20]. As a high-throughput computational approach [21], radiomics analysis combined with classical machine learning methods has been studied for glioma segmentation [17, 22]. The extracted radiomics features serve as computational biomarkers of image intensity texture and morphology descriptors, which are subsequently used by classifiers to distinguish glioma from normal tissue [23-27]. Recently, deep learning methods have been considered as a new computational approach for glioma segmentation [12, 28-30]. Deep neural networks can directly learn image features at multiple scales for segmentation tasks [31, 32], and the state-of-the-art deep learning models have achieved outstanding performances in glioma segmentation [12, 29, 30]. However, despite promising results in research works, potential clinical applications of deep learning models in glioma segmentation remain challenging [22, 33]. One major issue of currently available deep learning models is the lack of model explainability, i.e., the extent to which the internal mechanisms of a deep neural network can be explained in human terms from a clinical perspective. The explainability ensures that the networks are driven by (1) deep features that are appropriate for clinical practice and (2) decisions that are clinically defensible [34-36]. Without such model explainability, deep learning algorithms remain a 'black box' in implementation. The massive data computations in deep neural networks are beyond human logical and symbolic abilities for causality [37], which raises technical issues of deep learning model development for medical imaging applications. These issues include, but are not limited to, the utilization of model input ('*Do we need this as a part of the model?*'), the confidence of model results ('*Can I trust this run with some clues?*'), and feasibility of model generalization ('*How do I know if it works for this case?*'). Eventually, the absence of model explainability causes a lack of accountability and confidence in clinic application.

In this paper, we develop a deep neural network visualization method for multi-parametric MRI based glioma segmentation to enhance model explainability. Driven by the hypothesis that deep features extraction from multi-source input can be modeled as a spatiotemporal continuous process, we model a deep neural network's feature extraction process by an ordinary differential equation (ODE) [38]. Such a model (i.e., a neural ODE) can provide visualization effects of MRI image dynamics following their interactions with the deep neural network. A quantitative evaluator, Accumulated Contribution Curve (ACC), is designed to evaluate the image dynamics as a measure of each MRI modalities' utilization in glioma segmentation. Several comparison studies are included to verify the key MR modalities identified by ACC in glioma segmentation.

## 2. Materials and Methods

### A. Image Data

This work employed the Brain Tumor Segmentation Challenge (BraTS) 2020 dataset, a public-available large MRI image dataset from 369 subjects with glioblastoma (GBM/HGG) and low-grade glioma (LGG) [7, 12, 39]. Each subject has 4 standard MRI exams as a multi-parametric protocol: native T1 weighted (T1), contrast-enhanced T1-weighted (T1-Ce), T2-weighted (T2), and Fluid Attenuated Inversion Recovery (FLAIR). The ground-truth tumor segmentation from experienced neuro-radiologists comprises three non-overlapping subregions: contrast-enhanced tumor (ET), peritumoral edema (ED), and the necrotic and non-enhancing tumor core (NCR/NET). Following previous studies related to the BraTS dataset [7, 10], another three overlapped tumor regions (from small to large) are adopted in this work, namely enhancing tumor (ET), tumor core (TC), and whole tumor (WT). Figure 1 illustrates the relationship of ground-truth segmentation masks. All MRI images in the BraTS dataset have been processed with skull stripping and resampled to the isotropic 1mm spatial resolution.

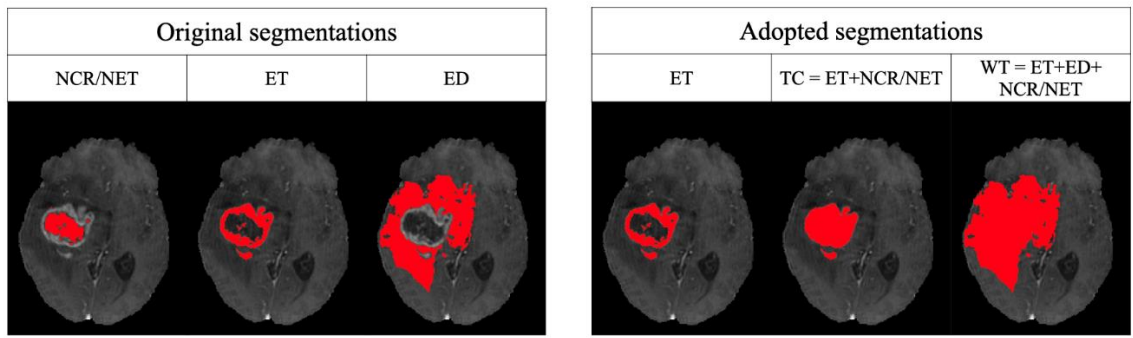

*Figure 1: The ground-truth tumor segmentation from BraTS 2020 dataset comprises the three non-overlapping subregions: contrast-enhanced tumor (ET), peritumoral edema (ED), and the necrotic and non-enhancing tumor core (NCR/NET). In this work, three overlapped tumor regions are adopted, namely enhancing tumor (ET), tumor core (TC), and whole tumor (WT).*

## B. Deep Learning Model Design

### B.1 Mathematical Modelling

We hypothesize that deep feature extraction by a deep neural network during image segmentation can be described mathematically as a spatiotemporal continuous process. As visualized in Figure 2, an image source before neural network input gradually evolves into a deep feature map towards binarized segmentation. Here, we refer to this derivable evolution as an image flow $\Phi$ [40]. When multiple image sources are available, i.e., multi-modality images for segmentation, each individual image source possesses an image flow during the deep feature extraction process.

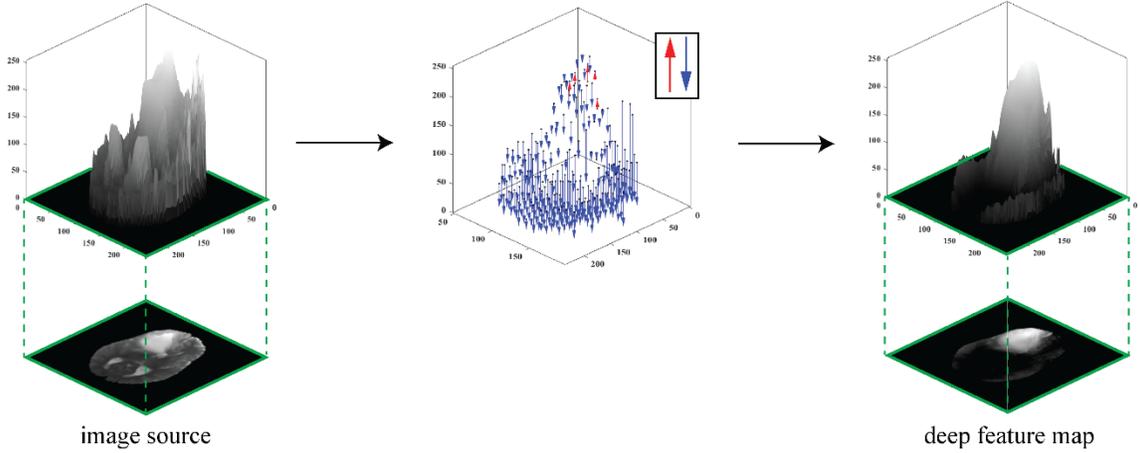

*Figure 2: The deep feature extraction by the deep neural network during image segmentation can be described as a continuous process. In such a process, an image source before neural network input gradually evolves into a deep feature map towards binarized segmentation. We refer to this derivable evolution as an image flow.*

Mathematically, let $\mathbf{M}$ be the collection of all $m$ available image sources, i.e., $\mathbf{M} = 1, 2, \ldots, m$. We define $t = 0$ and $t = 1$ are the initial stage and end stage of the modeled image flow $\Phi$, respectively, and $t$ is the transitional stage variable (i.e., 'time') between 0 and 1. Let $\mathbf{I}_0 = \{I_0^k\}_{k \in \mathbf{M}}$ is the input multi-source images, and $\mathbf{I}_1 = \{I_1^k\}_{k \in \mathbf{M}}$ is the collection of final deep features. We can formulate $\mathbf{I}_0 \rightarrow \mathbf{I}_1$ image flows $\mathbf{\Phi}$ ($\mathbf{\Phi} = \{\Phi_k\}_{k \in \mathbf{M}}$) as an ordinary differential equation (ODE) initial value problem:

$$\frac{d\mathbf{I}(t)}{dt} = f(\mathbf{I}(t), t, \theta), \quad \mathbf{I}(0) = \mathbf{I}_0 \qquad (1)$$

The hidden state $\mathbf{I}_t = \{I_t^k\}_{k \in \mathbf{M}}$ (i.e., the collection intermediate image frames within $\Phi$ at $t$) can be considered as the solution to the ODE initial value problem:

$$\mathbf{I}_t = \mathbf{I}_0 + \int_0^t f(\mathbf{I}(t), t, \theta) \mathrm{d}t. \tag{2}$$

As such, $\mathbf{I}_1 = \mathbf{I}(1)$ denotes the final deep feature map. Equation (1) is the core function of our mathematical formulation. The function $f$, parameterized by $\theta$, defines the continuous dynamics of the image flows $\Phi$. The hidden state $\mathbf{I}_t$ can be illustrated in the same manner as anatomical MRI images with explicit contrast, and $\Phi$ can be visualized via acquiring $\mathbf{I}_t$ at multiple stages. Such visualization provides visual clues of the MRI image evolution through the deep neural network, which serves as natural semantics to explain deep neural network behavior that can be understood by human experience [41-43].

*B.2 Neural ODE Structure*

The key task of our deep learning model design is to find $f$ in Equation (1). Based on the universal approximation theorem, $f$ can be approximated by a deep neural network that does not require explicit expression as in physical models [44, 45]. While $f$ does not require specific neural network architecture, in this work we focus on U-Net, which is a suitable and popular method for medical image segmentation [29]. The core design of U-Net architecture in this work is shown in Figure 3(A). Specifically, the structure consists of an encoding part and a decoding part. The encoding part is a convolutional network that consists of five convolutional blocks. Each block contains two convolutional layers followed by a rectified linear unit (ReLU) and a max pooling operation; in this process, the spatial dimension decreases while feature information increases. The decoding part combines the feature and spatial information through the corresponding five up-convolutions and concatenations with high-resolution features from the encoding part; in this process, the spatial dimension increases while feature information decreases. Such a U-shaped structure benefits feature extraction and preserves the structural integrity of the image [46].

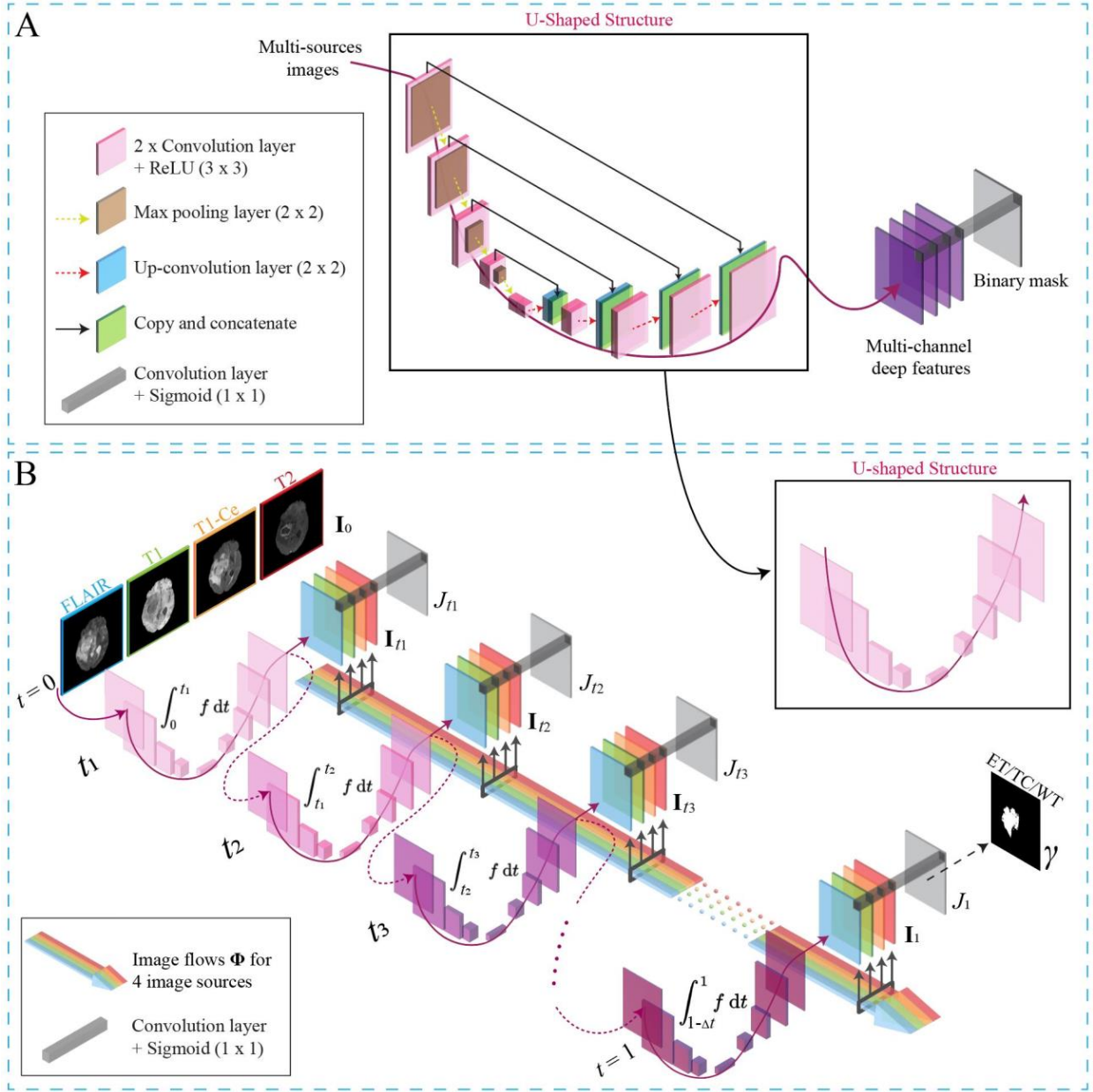

*Figure 3. (A) The adopted U-Net architecture is featured by an encoding path and a decoding path. (B) The conceptional design of the proposed neural ODE model architecture. The model describes the image flow from four MRI images to four deep feature maps extracted by the U-Net in (A). $I_t$ is the hidden state within image flow $\Phi$ at stage t. $J_t$ is the corresponding intermediate segmentation frame along with $I_t$. $\gamma$ is the final segmentation result (ET/TC/WT).*

In this work, we parametrize the function of $f$ by the U-Net shown in Figure 3 (A), which takes 4 MR modalities (FLAIR, T1, T1-Ce, and T2) as input. Taking together, we propose a neural ODE design based on Equation (1) and Figure 2 to visualize mp-MRI based glioma segmentation by U-Net. Figure 3 (B) summarizes this design. Formally, this design models the image flow from four MRI images to four deep feature maps extracted through the U-Net's U-shaped structure, where $\mathbf{I}_0$ represents MRI images from 4 modalities (FLAIR, T1, T1-Ce, and T2), and the blue, green, orange, and red bands demonstrate the corresponding image flow $\Phi$, respectively. The dynamics of the flow can be interpreted by integrating the $f$ (U-shaped structure) from 0 to different $t$. At each state $t$, the hidden state $\mathbf{I}_t$ is represented as a 4th order tensor object capturing the information of the four images, where the colors correspond to four different image modalities. The gradient color of the U-shaped structure represents increasing integration interval on $t$ dimension.

To reach final segmentation results as a binarized mask, a convolutional layer follows the final deep feature maps at $t=1$. This convolutional layer assigns a binary label to each pixel by performing weighted summation on the final deep feature maps $\mathbf{I}_1$:

$$\gamma = \sigma\left(\sum_{k=1}^{m} w^k I_1^k\right). \tag{3}$$

where $\sigma$ is the sigmoid function, and $w^k$ is the learnable weight of image modality $k$ in the convolutional layer. Thus, $m$ (=4 in this study) deep feature maps are constructed to one binarized segmentation mask as in Figure 3 (B). For a trained model, the weight $w^k$ is known, so $w^k$ can be used to reconstruct intermediate segmentation frame $J_t$ along with the hidden state $\mathbf{I}_t$:

$$J_t = \sigma\left(\sum_{k=1}^{m} w^k I_t^k\right). \tag{4}$$

Evidently, when $t = 1$, we reach the final segmentation result $J_t = \gamma$. The collection of $J_t$ can also be visualized to illustrate the segmentation formation.

By providing the data pair $(\mathbf{I}_0, \gamma)$, we train the model in Figure 3 (B) as a supervised learning problem. The forward problem in Equation (2) can be solved by standard ODE solvers. During the training phase, the gradients of the output $\mathbf{I}_1$ with respect to the input $\mathbf{I}_0$ and the parameter $\theta$ can be obtained using the adjoint sensitivity method; this includes solving an additional ODE backward in stage dimension and thus invoking another ODE solver in the backward pass. Once the gradient is obtained, the standard gradient-based optimization can be applied. For technical details we refer to the mathematical works in [38, 40, 47-49].

*B.3 Quantitative Model Analysis*

While the hidden state $\mathbf{I}_t$ and segmentation frame $J_t$ provide visual clues of U-Net behavior in segmentation, the image-based illustration may be insufficient to quantify the importance of individual input (i.e., utilization) towards segmentation results. Specifically, micro-regional changes of $\mathbf{I}_t$ after interactions with U-Net may lead to prominent changes of $J_t$, while drastic changes of $\mathbf{I}_t$ morphology may not contribute to the effective progress of $J_t$. We hereby propose a quantitative evaluation of $\mathbf{I}_t$ to further explain the roles of each image flow $\Phi$ in multi-modality image segmentation.

Based on Equation (3), the sigmoid function outcome depends on the polarity of weighted deep features summation. Although the polarity in Equation (3) cannot be directly interpreted ($w^k$ comes from an unnormalized operation), the comparison of weighted deep features can indicates the contributions of different image modalities: for each MRI modality $k$ at stage $t$, if the weighted hidden state $w^k I_t^k$ is larger than the weighted sum of the other modalities, we argue MRI modality $k$ has a positive and leading contribution to the segmentation at $t$. We describe this as binarized contribution factor (BCF) $\boldsymbol{\eta}_t = \{\eta_t^k\}_{k \in \mathbf{M}}$:

$$\eta_t^k = \begin{cases} 1, & \text{if } \left( w^k I_t^k - \sum_{n \neq k} w^n I_t^n \right) > 0 \\ 0, & \text{otherwise} \end{cases} \tag{5}$$

where $k \in \mathbf{M}, n \in \mathbf{M}$. As defined, BCF $\eta_t^k$ remains in the same dimension as a single MRI image source, and it can be visualized along with $\mathbf{I}_t$ and $J_t$. To quantify the spatiotemporal heterogeneity of $\eta_t^k$, for each MRI modality $k$, we compare $\eta_t^k$ and final segmentation results $\gamma$ and define accumulated contribution factor as:

$$\mu_t^k = \frac{|\eta_t^k \cap \gamma|}{|\eta_t^k \cup \gamma|},  \tag{6}$$

where ∩ and ∪ are intersection and union operations of the Boolean image, respectively, and $\mu_t^k$ ranges from 0 to 1. The collection of $\mu_t^k$, namely accumulated contribution curve (ACC), describes the dynamic properties of an image modality's role in multi-modality image segmentation: when ACC has a positive slope (as a function of *t*), the deep learning model tends to extract more useful deep features from the studied MRI modality towards a positive contribution of final segmentation results. An ACC with a steeper slope and higher value when approaching *t* = 1 suggests that the studied MRI modality is more important within the multi-parametric imaging protocol for the segmentation task.

## C. Comparison Studies

All calculations in this work were carried out in Python 3.7 with 8 Core Intel Xeon Silver 4112 CPU @ 2.6 GHz, 32 GB RAM, and Nvidia Quadro P5000 Graphic Card. We implemented the neural ODE deep learning model in Figure 3 (B) with TorchDyn Library [50, 51]. A total of three independent models were trained to achieve segmentation results of ET, TC, and WT regions, respectively. A total of 24422 samples as 4-channel 2D MRI images were randomly divided into the training set and test set in the ratio of 8:2 in the patient assignment. The Adam optimizer with the initial learning rate of $10^{-3}$ was applied. The standard $4^{th}$ order Runge-Kutta ODE solver was adopted as an ODE solving engine with $10^{-3}$ tolerance, and the adjoint sensitivity method was employed to optimize parameters in the U-shape structure. The parameters in the last 1x1 convolutional layer were also updated in the backpropagation. In the model trained for each segment region, $\mathbf{I}_t$, $J_t$, and $\mathbf{\eta}_t$ results were saved at a step size of 0.01 from $t = 0$ to $t = 1$, and ACCs were calculated based on final segmentation results $\gamma$.

Based on the qualitative inspection of $\mathbf{I}_t$, $J_t$ and $\mathbf{\eta}_t$, as well as quantitative analysis of ACCs, we identified key MRI modalities of each segmentation region. We then evaluated the following two studies:

1) Baseline segmentation results (ET/TC/WT) from the basic U-Net model. i.e., the U-Net model in Figure 3(A) that uses all 4 available MRI modalities as input.
2) Segmentation results (ET/TC/WT) from a modified U-Net model. i.e., a U-Net model with the same design as in the baseline model but has reduced input dimensions using only the identified key MRI modalities from neural ODE results.

In both studies, the same training and test sets were used as in the neural ODE model. The loss function was binary cross-entropy, and the initial learning rate in the Adam optimizer was $10^{-3}$. Ten-fold cross-validation within the training set was employed to evaluate the segmentation results objectively. Following the previous BraTS related studies [50], we included Dice similarity index, accuracy, sensitivity, and specificity evaluation metrics. For a successful neural ODE design and analysis, we expect minimal difference in segmentation results between these two studies. Wilcoxon signed-rank test on Dice index with significance level 0.05 was adopted.

## 3. Results

Figure 4(A) provides an example of 4 image flows **Φ** in ET segmentation. The 1st row contains 4 input MRI images $\mathbf{I}_0$ at *t* = 0, and the last row contains 4 final deep feature maps $\mathbf{I}_1$. The 2nd to 4th rows are the visualized hidden states $\mathbf{I}_t$ at *t*=0.3, 0.5, and 0.8, respectively. As illustrated, the neural ODE model successfully renders 4 MRI image evolutions in classic MRI contrasts. Based on the visual inspection, while T1 and T2 show prominent morphological changes, FLAIR and T1-Ce demonstrate subtle contrast changes limited within high-intensity regions. The intermediate segmentation frames $J_t$ at each stage are shown in Figure 4(B) along with $\mathbf{I}_t$. The formation process of the binary segmentation from the whole brain to localized ring-shape result is clearly illustrated, and the final segmentation result $\gamma$ at *t* = 1 is very similar to the ground truth. Figure 4(C) shows the corresponding BCF **η** of 4 MRI image sources. The highlighted region of $\eta_t^k$ represents the spatial regions with a positive contribution towards the segmentation results at *t*. The BCF $\eta_t^{\text{T1-Ce}}$ demonstrates higher morphological similarity to the final segmentation results as $t \to 1$, and this observation suggests that T1-Ce played a leading role in ET segmentation. The BCF $\eta_t^{\text{FLAIR}}$ shows similarly and yet less prominent visual results as in $\eta_t^{\text{T1-Ce}}$, and the role of FLAIR may need further investigation in more quantitative studies.

Figures 5-6 present ET and WT results in the same Figure 4 manner. Essentially, the role of each MRI modality varies significantly for different segmentation targets. In the TC segmentation, only T1 and T2 image flows shows a noticeable change, and neither of them highlighted TC region. In the WT segmentation, all four modalities presents certain degrees of morphological changes in their image flows. Based on the visual clues of BCF, $\eta_t^{\text{T1-Ce}}$ again demonstrates highest morphological similarity to $\gamma$ in TC segmentation, and both $\eta_t^{\text{FLAIR}}$ and $\eta_t^{\text{T2}}$ demonstrat high similarity in WT segmentation. These observations suggest that T1-Ce played a leading role in TC segmentation, and FLAIR and T2 are both key modalities in WT segmentation. Alternative illustrations of Figures 4-6 as animations can be found [here](#) as well as in Supplementary Data.

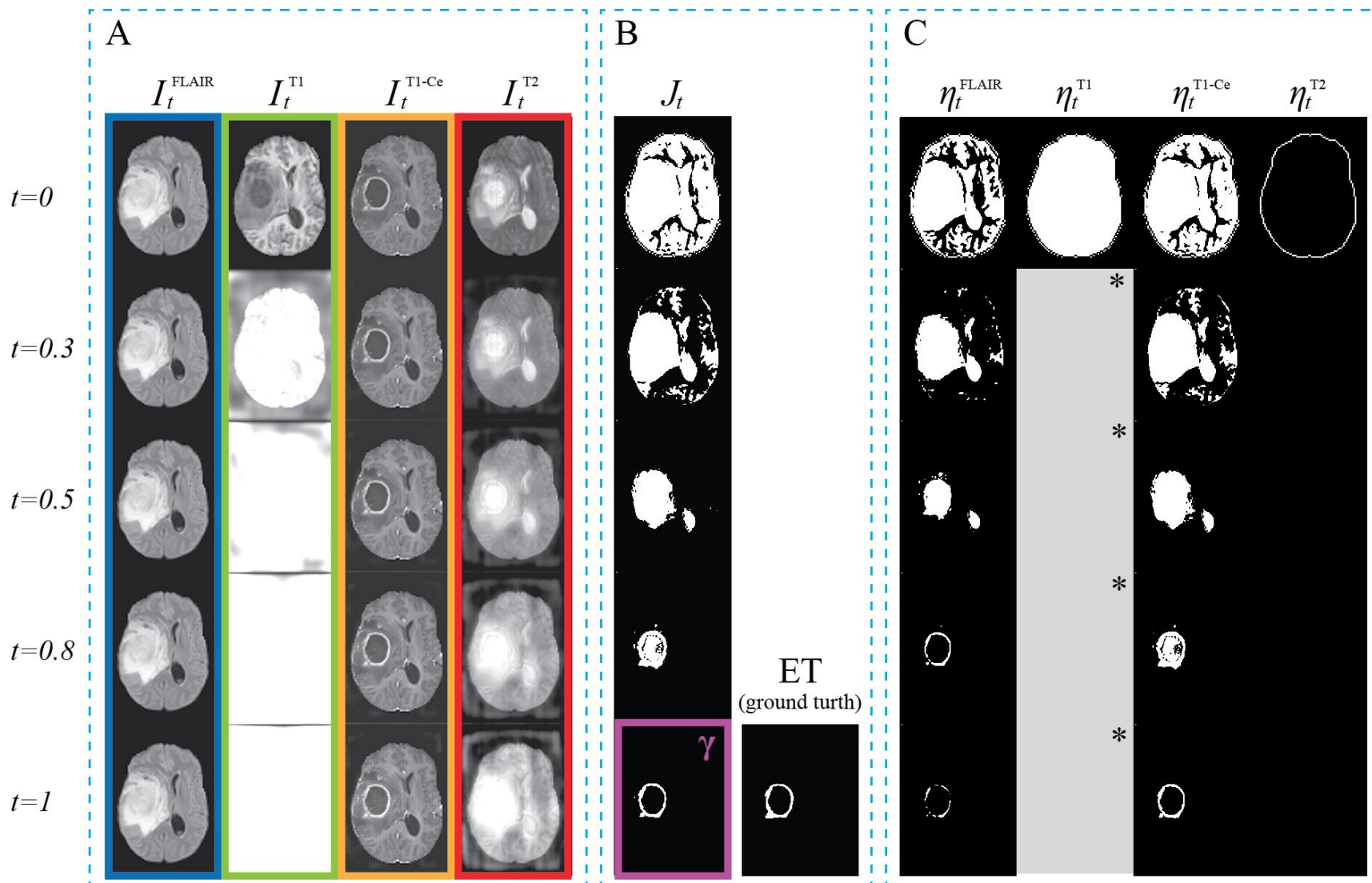

*Figure 4. (A) 4 image flows $\Phi$ in ET segmentation based on the neural ODE model. (B) The corresponding intermediate segmentation frames $J_t$ and the ground truth ET segmentation. (C) The corresponding binarized contribution factor (BCF) $\eta$. The maker "*" indicates unity value in the whole map without spatial patterns.*

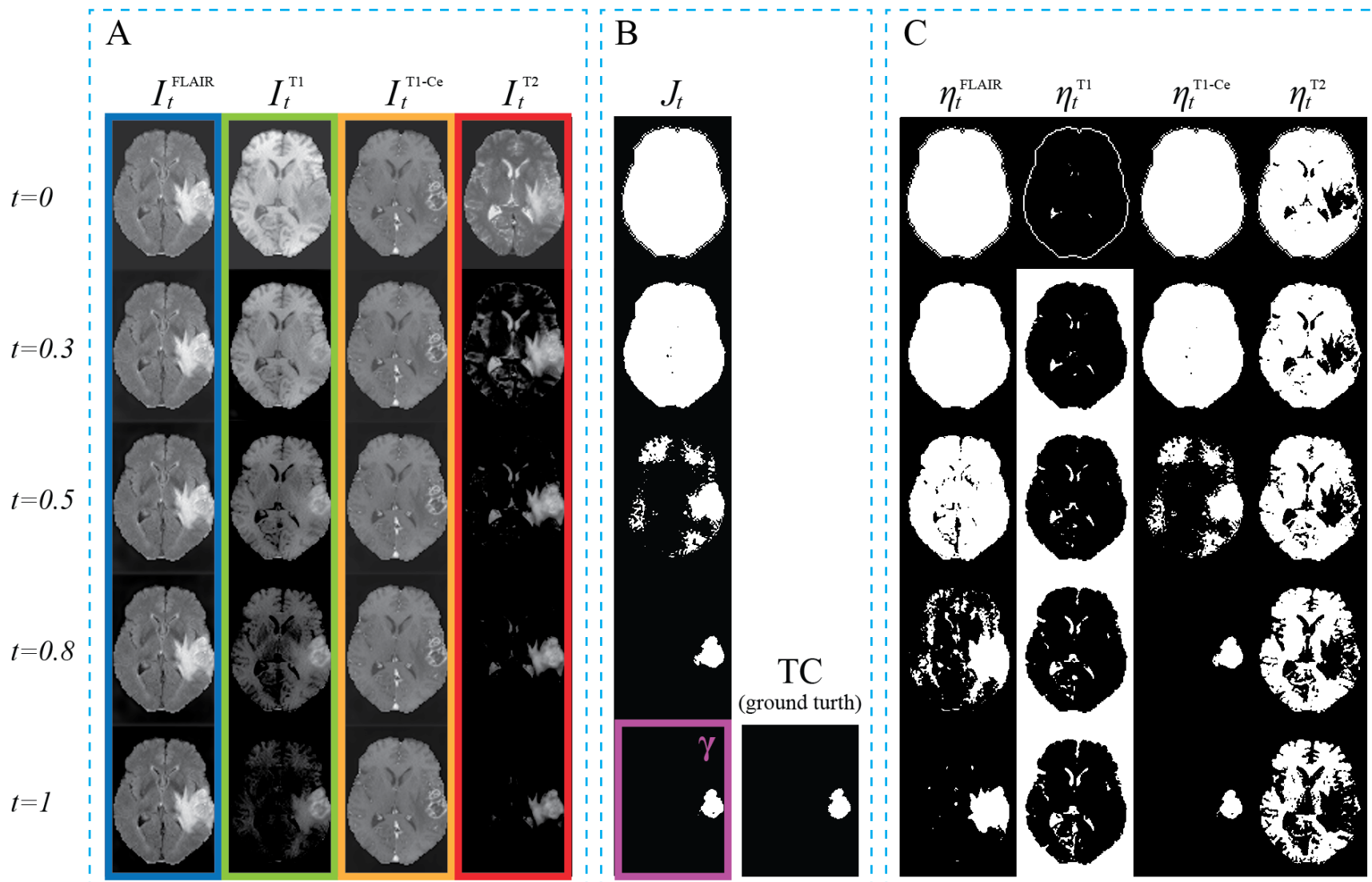

*Figure 5. (A)-(C) 4 image flows **Φ**, the corresponding intermediate segmentation frames $J_t$, and the corresponding binarized contribution factor (BCF) **η** in TC segmentation based on the neural ODE model, respectively.*

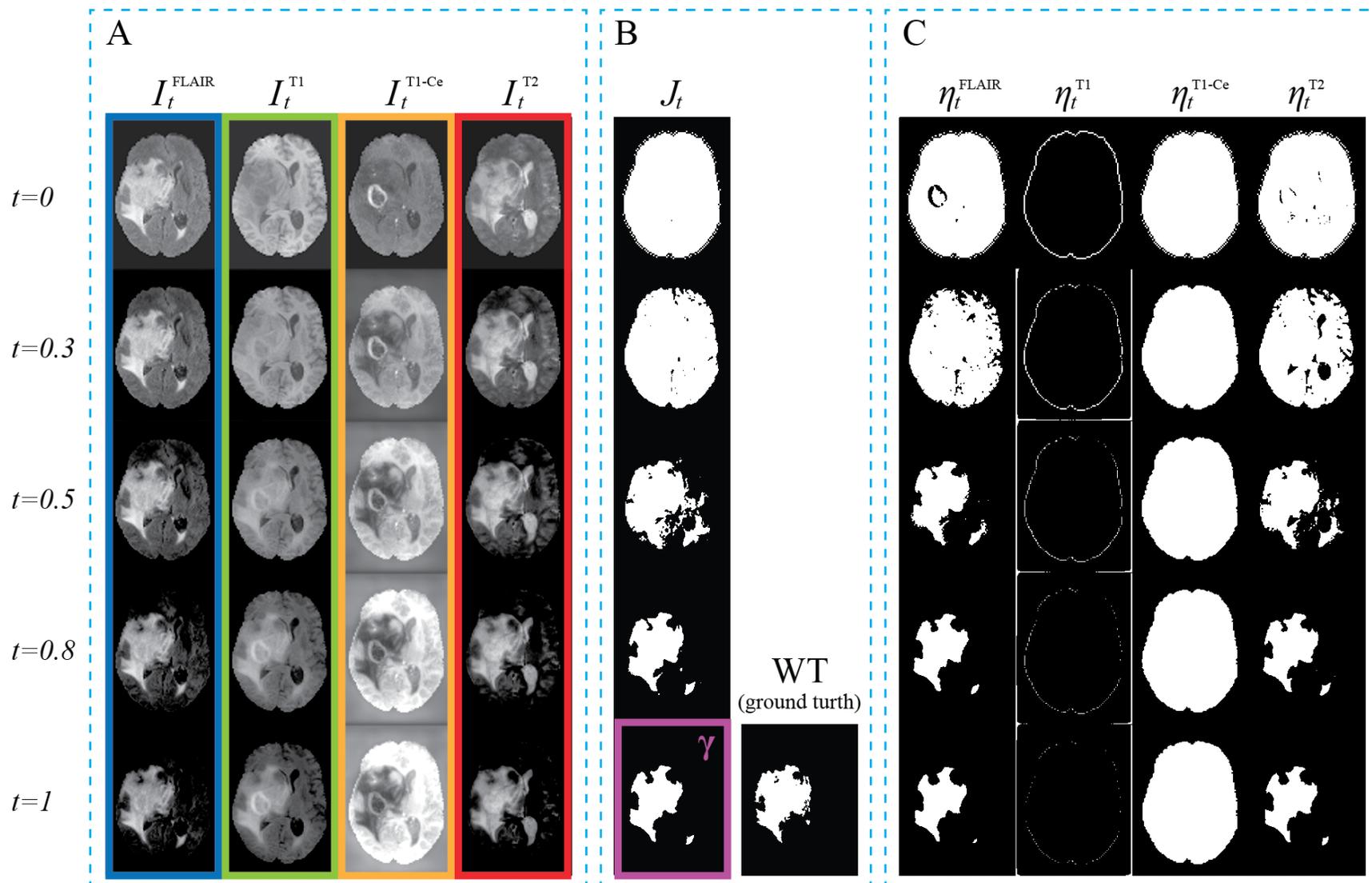

*Figure 6. (A)-(C) 4 image flows **Φ**, the corresponding intermediate segmentation frames $J_t$, and the corresponding binarized contribution factor (BCF) **η** in WT segmentation based on the neural ODE model, respectively.*

Figure 7(A)-(C) present the ACC results ($\mu_t^k$) as a function of $t$ for ET, TC, and WT segmentation, respectively. Each ACC is shown as a shaded plot, where the solid line represents the mean ACC of all test images, and the shaded area indicates the standard deviation from all test images. In general, the results in Figure 7 agree with visual interpretations of Figures 4-6. Here, T1-Ce presents the highest positive ACC slope and very high value (close to 1) for $t \rightarrow 1$ in both ET and TC segmentations. While FLAIR ACC in TC segmentation shows a negligible increase from $t = 0$ to $t = 1$, FLAIR in ET segmentation initially shows a noticeable increase in ACC followed by a decrease after attaining its peak value at around 0.3. At $t = 1$, $\mu_1^{\text{FLAIR}} < 0.2$, and thus FLAIR was not considered as a key modality candidate compared to T1-Ce. As such, T1-Ce can be considered as the most important input source to the model to achieve accurate ET and TC segmentation. In WT segmentation, both FLAIR and T2 ACC results demonstrate a big initial slope and high values at $t = 1$; in contrast, T1 and T1-Ce have steady ACC with low values. Thus, FLAIR and T2 can be recognized as the key modalities for WT segmentation.

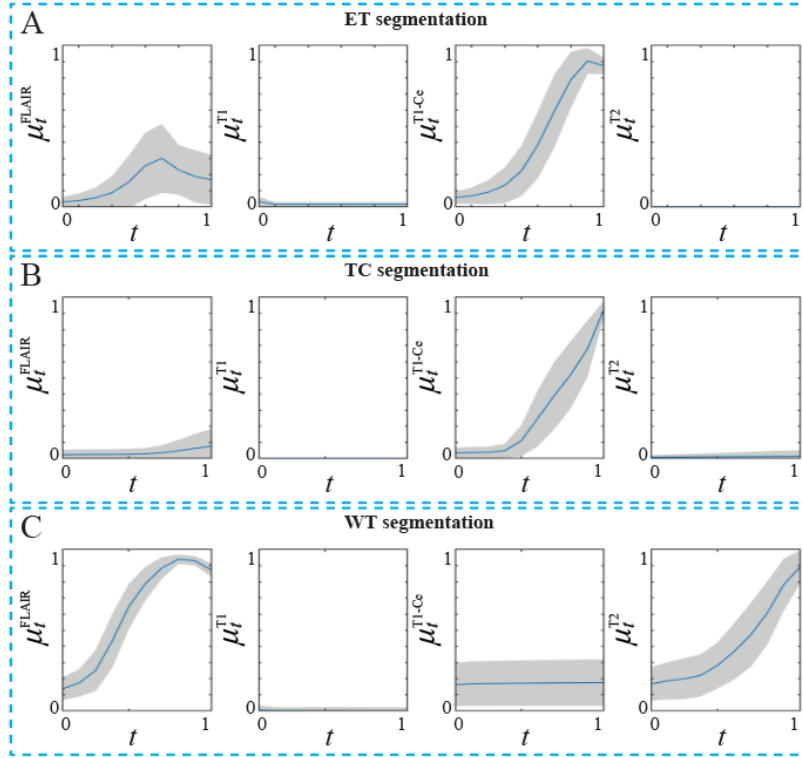

*Figure 7. (A)-(C) presents the $\mu_t^k$ as a function of t for ET, TC, and WT segmentation, respectively. The collection of $\mu_t^k$ from t=0 to t=1 is named accumulated contribution curve (ACC). The solid blue lines represent the mean ACC across all test images, and the shaded areas indicate the corresponding standard deviation.*

Table I summarizes the ten-fold cross-validation segmentation results (mean ± standard deviation) in the comparison studies. The results from all four MRI modalities are the baseline results. As the main numeric evaluator, the Dice coefficients from segmentation results using neural ODE selected modalities are very close to baseline results without statistically significant differences. The other three evaluators also show minimal numeric differences. In contrast, when using complementary MRI modalities, i.e., ones not selected by the neural ODE model, the Dice coefficients are much lower than baseline results with significant differences. In addition, sensitivity and accuracy results are compromised when using complementary modalities, but specificity seems insensitive to MRI modality selection. Table II summarizes the segmentation of the neural ODE model on the test set. The achieved quantitative results in all three segmentation regions are very close to the baseline results in Table I.

Table I. Ten-fold cross-validation segmentation results in the comparative study (mean ± standard deviation) using U-net. The results from all four MRI modalities are the baseline results. "*" indicates a statistically significant difference compared with baseline results

|   |   | Inputs | | |
|---|---|---|---|---|
|   |   | 4 Modalities | T1-Ce | FLAIR+T1+T2 |
| ET | Dice | 0.784±0.008 | 0.775±0.007 | 0.295±0.016* |
|    | Accuracy (%) | 99.537±0.015 | 99.032±0.031 | 98.417±0.045* |
|    | Sensitivity | 0.919±0.018 | 0.949±0.016 | 0.875±0.014* |
|    | Specificity | 0.999±0.001 | 0.999±0.001 | 0.998±0.000 |
|   |   | 4 Modalities | T1-Ce | FLAIR+T1+T2 |
| TC | Dice | 0.760±0.014 | 0.758±0.006 | 0.042±0.010* |
|    | Accuracy (%) | 99.005±0.063 | 99.489±0.008 | 96.929±0.132* |
|    | Sensitivity | 0.949±0.016 | 0.909±0.015 | 0.031±0.005* |
|    | Specificity | 0.999±0.000 | 0.999±0.001 | 0.997±0.001* |
|   |   | 4 Modalities | FLAIR+T2 | T1+T1CE |
| WT | Dice | 0.841±0.007 | 0.837±0.004 | 0.691±0.010* |
|    | Accuracy (%) | 98.962±0.033 | 98.922±0.026 | 98.124±0.061 |
|    | Sensitivity | 0.957±0.011 | 0.962±0.008 | 0.943±0.007* |
|    | Specificity | 0.999±0.001 | 0.999±0.001 | 0.998±0.000 |

Table II. Segmentation results of the proposed neural ODE model.

|   | ET | TC | WT |
|---|---|---|---|
| Dice | 0.775 | 0.845 | 0.773 |
| Accuracy (%) | 99.649 | 99.425 | 99.383 |
| Sensitivity | 0.774 | 0.839 | 0.762 |
| Specificity | 0.998 | 0.997 | 0.998 |

## 4. Discussion

In this work, we demonstrated a deep learning neural ODE design that models the multi-parametric MRI-based glioma segmentation as a continuous process. One of the key innovations in this work is the spatiotemporal continuous segmentation modeling, which was conceptualized as an ODE that governs the deep feature extraction process. Here, by explicitly specifying the ODE as a U-Net architecture, the deep feature maps are constrained in the same contrast as anatomical images, and thus image transitions after interactions with U-Net can be visually inspected as image flows. The contrast change in image flows represents the pixel-level interaction of the network to each input MRI image. One interesting finding of our work is that a strong interaction between an MRI image and the deep neural network does not guarantee a 'positive' contribution to the segmentation. For example, $I_t^{T1}$ in Figure 4(A) gradually evolves to a unity map without any spatial patterns. As a result, the BCF and ACC are designed to further analyze spatiotemporal heterogeneity of U-Net behavior. The current results suggest that deep learning-based glioma segmentation can be done using a multi-parametric protocol with a reduced number of MRI modalities. This leads to improvements of deep learning model design in both technical and clinical aspects: in a deep learning model, when using excessive data source as neural network input without optimization, the model training could be extremely slow due to unnecessarily complex architecture design, and the results may suffer potential over-fitting problem, particularly in medical imaging problems with smaller dataset sizes [52]. Still, the reduced input design means reduced procedures and exams of clinical data acquisition, which may benefit the study protocol design and potential clinical practice. It is mandatory to evaluate the role of multiple image sources within a single deep learning model. In fact, previous studies investigated the role of each image source via dropout comparison studies, i.e., compare two deep learning models' results with/without a specific image source as input [53]. However, such comparison relies on the results from two models with different dimension designs and training histories, and this not-equal comparison of final results adds limited value to explaining the role of the investigated image. The developed neural ODE model addresses such explainability issues and provides a powerful tool to evaluate the utilization of model input in medical image tasks.

In addition to the input utilization issue, the developed model also helps to address the results confidence issues. Based on the $J_t$ results in Figures (4)-(6), the formation process of the binary segmentation mask can be visualized. Our results successfully illustrated how each segmentation mask was learned by U-Net through transitions from the whole brain to the localized tumor region. The transitions from global to local content and from coarse to fine structure are consistent with human intuition. In general, T1 scans highlight fat tissues within the body. T1-Ce indicates the active lesions by enhancing the blood-brain barrier breakdown region. T2 and FLAIR scans highlights the tissue water of the edema [7, 54]. Our ACC analysis identified that T1-Ce is the only key modality in ET and TC segmentations, while both FLAIR and T2 are key modalities in WT segmentation. The identified key MR modalities are consistent with clinicians' experience in radiography reading. These consistencies suggest that the deep features extraction in U-Net ensembles the human recognition of glioma. Therefore, the developed model may help clinicians understand the deep neural network behavior and raise the confidence of deep learning applications in the clinic. Additionally, the $J_t$ and $\eta_t$ results may serve as visual aids to help estimate quantification uncertainty for the sake of model generalization [55-57]. When the neural ODE model processes a new case, the intermediate results can be compared with previous cases to see if they follow the patterns in previous cases; thus, the final segmentation results will be more convincing with support from additional data. The developed neural ODE model can also be adapted for other tasks by replacing U-Net with different deep neural network architectures.

The current neural ODE model was designed and implemented in a 2D manner for a better image visualization effect. Additionally, such 2D-based implementation is consistent with the manual glioma delineation process, during which radiologists focus on 2D axial view during contouring. The current results in Figures (4)-(6) are suitable following clinical conventions. In theory, specifying the ODE in 3D space for 3D segmentation is plausible; however, given the fact that ground-truth data did not come from direct 3D contouring, the 3D implementation does not guarantee improved segmentation results. In addition, the 3D neural ODE model may only use 369 independent samples, while the current 2D implementation utilizes >20k images as independent samples. Since the transferred learning scheme cannot be adopted in the neural ODE training, a limited 369 sample size is too small to reach model convergence during training. In addition, neural ODE requires significant computational resources. By keeping the same training configurations as the current 2D-based design, the 3D-based calculation is estimated to cost more than one terabyte of RAM and GPU memory, which is infeasible using state-of-the-art hardware. Future collaboration with high-performance computing (HPC) core is desired to investigate such a huge computation task.

As a feasible study, the current ACC analysis for key MRI modality is based on curve shape comparison. The adopted curve feature criteria may be sufficient for 4 input candidates. In future works, when more input channels are modeled in the neural ODE, more quantitative ACC features with sophisticated statistical modeling, such as the clustering method, may be required to identify useful input channels [58, 59]. Another future research direction is asynchronous deep feature extraction. When ACC shows a nonmonotonic shape (such as FLAIR ACC in Figure 7(A)), deep features from an image source may exhibit the highest contribution to segmentation results at a certain intermediate state instead of the final state. It would be meaningful to investigate if deep features at different stages can be connected for segmentation and/or classification purpose. To realize such works, deep neural network architecture may need further optimization to allow such feature connections with open-dimension design.

## 5. Conclusion

In this work, we developed a neural ODE model to realize image evolution visualization for deep learning-based image segmentation. In the multi-parametric MRI-based glioma segmentation study, the developed model successfully identified key MR modalities via both qualitative visual clues and quantitative image dynamics analysis. The developed neural ODE model enhances the explainability of U-Net deep learning models, and the presented methodology can be generalized to other deep learning models for improved clinical applications.

# Figure Captions

Figure 1: The ground-truth tumor segmentation from BraTS 2020 dataset comprises the three non-overlapping subregions: contrast-enhanced tumor (ET), peritumoral edema (ED), and the necrotic and non-enhancing tumor core (NCR/NET). In this work, three overlapped tumor regions are adopted, namely enhancing tumor (ET), tumor core (TC), and whole tumor (WT).

Figure 2: The deep feature extraction by the deep neural network during image segmentation can be described as a continuous process. In such a process, an image source before neural network input gradually evolves to a deep feature map towards binarized segmentation. We refer to this derivate evolution as an image flow.

Figure 3. (A) The adopted U-Net architecture is featured by an encoding path and a decoding path. (B) The conceptional design of the proposed neural ODE model architecture. The model describes the image flow from four MRI images to four deep feature maps extracted by the U-Net in (A). $I_t$ is the hidden state within image flow $\Phi$ at stage $t$. $J_t$ is the corresponding intermediate segmentation frame along with $I_t$. $\gamma$ is the final segmentation result (ET/TC/WT).

Figure 4. (A) 4 image flows $\boldsymbol{\Phi}$ in ET segmentation based on the neural ODE model. (B) The corresponding intermediate segmentation frames $J_t$ and the ground truth ET segmentation. (C) The corresponding binarized contribution factor (BCF) $\boldsymbol{\eta}$. The maker "*" indicates unity value in the whole map without spatial patterns.

Figure 5. (A)-(C) 4 image flows $\boldsymbol{\Phi}$, the corresponding intermediate segmentation frames $J_t$, and the corresponding binarized contribution factor (BCF) $\boldsymbol{\eta}$ in TC segmentation based on the neural ODE model, respectively.

Figure 6. (A)-(C) 4 image flows $\boldsymbol{\Phi}$, the corresponding intermediate segmentation frames $J_t$, and the corresponding binarized contribution factor (BCF) $\boldsymbol{\eta}$ in WT segmentation based on the neural ODE model, respectively.

Figure 7. (A)-(C) presents the $\mu_t^k$ as a function of t for ET, TC, and WT segmentation, respectively. The collection of $\mu_t^k$ from $t=0$ to $t=1$ is named accumulated contribution curve (ACC). The solid

blue lines represent the mean ACC across all test images, and the shaded areas indicate the corresponding standard deviation.